\documentclass[12pt]{article}

\usepackage{amsfonts,amssymb,bm,cite,amsmath}
\usepackage[dvipdfmx]{graphicx}
\usepackage{array, booktabs}
\usepackage{braket}
\usepackage[title]{appendix}

\newcommand {\beq}{\begin{equation}}
\newcommand {\eeq}{\end{equation}}
\newcommand {\beqa}{\begin{eqnarray}}
\newcommand {\eeqa}{\end{eqnarray}}
\newcommand {\n}{\nonumber \\}

\newcommand {\del}{\partial}

\begin{document}

\setlength{\oddsidemargin}{0cm}
\setlength{\baselineskip}{7mm}

%%%%%%%%%%%%%%%%%%%%%%%%%%%%%%%%%%%%%%%%%%%%%%%%%%
 \titlepage
%%%%%%%%%%%%%%%%%%%%%%%%%%%%%%%%%%%%%%%%%%%%%%%%%%

 \begin{titlepage}
 \renewcommand{\thefootnote}{\fnsymbol{footnote}}
 \begin{normalsize}
 \begin{flushright}
 \begin{tabular}{l}
December 2021
 \end{tabular}
 \end{flushright}
 \end{normalsize}

 ~~\\

 \vspace*{0cm}
     \begin{Large}
        \begin{center}
          {A geometrical representation of the quantum information metric \\ in the gauge/gravity correspondence}
        \end{center}
     \end{Large}
 \vspace{1cm}

 \begin{center}
            Asato T{\sc suchiya}\footnote
             {
 e-mail address :
 tsuchiya.asato@shizuoka.ac.jp}
            and
            Kazushi Y{\sc amashiro}\footnote
            {
 e-mail address : yamashiro.kazushi.17@shizuoka.ac.jp}\\
       \vspace{1.5cm}

 % $^{1)}$
 {\it Department of Physics, Shizuoka University}\\
                {\it 836 Ohya, Suruga-ku, Shizuoka 422-8529, Japan}\\
          \vspace{0.5cm}
 %        $^{2)}$
  {\it Graduate School of Science and Technology, Shizuoka University}\\
                  {\it 836 Ohya, Suruga-ku, Shizuoka 422-8529, Japan}
%                {\it 3-5-1 Johoku, Naka-ku, Hamamatsu 432-8011, Japan}

 \end{center}

 \vspace{3cm}

 \begin{abstract}
 \noindent
We study a geometrical representation of the quantum information metric 
in the gauge/gravity correspondence. We consider the quantum information
metric that measures the distance between the ground states of 
two theories 
on the field theory side, one of which is obtained by perturbing the other.
We show that the information metric is represented by a back reaction to 
the volume of a codimension-2 surface on
the gravity side if the unperturbed field theory possesses the Poincare symmetry.
% We study how information geometry is described by bulk geometry in the gauge/gravity correspondence. We consider a quantum information metric that measures the distance between the ground states of a CFT and a theory obtained by perturbing the CFT.
% We find a universal formula that represents the quantum information metric
% in terms of back reaction to the AdS bulk geometry.
 \end{abstract}
 \vfill
 \end{titlepage}

%%%%%%%%%%%%%%%%%%%%%%%%%%%%%%%%%%%%%%%%%%%%%%%%%%%%%%%%%%%%%%%%%%%%%%%%%%

\vfil\eject

\setcounter{footnote}{0}

%%%%%%%%%%%%%%%%%%%%%%%%%%%%%%%%%%%%%%%%%%%%%%%%%%%%%
\section{Introduction}
\setcounter{equation}{0}
\renewcommand{\thefootnote}{\arabic{footnote}}
It has recently been recognized that 
the quantum information theory seems to play a crucial role\cite{Ryu:2006bv}
in gaining deep understanding of the gauge/gravity correspondence\cite{Maldacena:1997re,Gubser:1998bc,Witten:1998qj}. For instance, the connection of
the quantum information metric with the bulk geometry has been investigated in \cite{MIyaji:2015mia,Bak:2015jxd,Trivella:2016brw,Chen:2018vkw,Karar:2019wjb,Nozaki:2012zj,
Lashkari:2015hha,Aoki:2017bru,May:2018tir,Tsuchiya:2020drh}.

In \cite{Tsuchiya:2020drh}, 
we found a universal formula that represents the quantum information metric
in terms of a back reaction to a geometrical quantity in the bulk.
We considered a CFT and a theory obtained by perturbing the CFT by a primary operator,
and calculated the quantum information metric that measures the distance between 
the ground states
of the two theories. We showed that the quantum information metric is represented by 
a back reaction to the volume of a codimension-2 surface. This is universal in the sense that it holds
for perturbations by scalar, vector and tensor operators.

In this letter, we push forward with the above project. 
We show that the above formula holds also for
a field theory, which is not necessarily
a CFT and has a gravity dual whose background geometry is not necessarily AdS.
We introduce a covariant calculation on the gravity side, which allows us to derive a condition 
that must be satisfied by the dual geometry in order that the quantum information metric can be 
represented by the back reaction to  the volume of the codimension-2 surface. 
We find that the condition implies that the original field theory possesses the Poincare invariance.

%%%%%%%%%%%%%%%%%%%%%%%%%%%%%%%%%%%%%%%%%%%%%%%%%%%%%

%%%%%%%%%%%%%%%%%%%%%%%%%%%%%%%%%%%%%%%%%%%%%%%%%%%%%
\section{On-shell action and Einstein equation}
\setcounter{equation}{0}
\renewcommand{\thefootnote}{\arabic{footnote}}
%%%%%%%%%%%%%%%%%%%%%%%%%%%%%%%%%%%%%%%%%%%%%%%%%%%%%

In this section, we consider a back reaction caused by a scalar field in  a background geometry
in the bulk.
The background geometry is supposed to be dual to a field theory on the boundary
which is perturbed by an operator
corresponding to the scalar field in the bulk.
We represent the on-shell action for the scalar field
in terms of the back reaction to the background geometry.
As seen in the next section, the information metric
that measures the distance between the ground states of the original and perturbed field theories
corresponds to the on-shell action for the scalar field.
Thus, we eventually obtain a formula that represents the information metric in terms of
the back reaction to the background geometry.

The coordinates in the $(d+1)$-dimensional bulk spacetime are denoted by
$x^{\mu} = (z, x^i)=(z,\tau,x^a)$ where $i = 0,\ldots, d-1$, $a= 1,...,d-1$ and $\tau=x^0$. 
The spacetime metric  takes the form
\begin{align}
  ds^2 = g_{\mu\nu} dx^{\mu}dx^{\nu} = g_{zz} dz^2 + g_{ij} dx^i dx^j \ ,
  \label{metric}
\end{align}
where $g_{\mu\nu}$ is expanded around a background $\hat{g}_{\mu\nu} $ as
\begin{align}
g_{\mu\nu} = \hat{g}_{\mu\nu} + \tilde{g}_{\mu\nu}
\label{expansion of metric}
\end{align} 
with
$ \tilde{g}_{\mu\nu}$ a perturbation.
A gauge condition $\hat{g}_{zi}=\tilde{g}_{zi}=\tilde{g}_{zz} = 0$ is imposed.
We assume that 
\begin{align}
\del_i \hat{g}_{\mu\nu} = 0 
\label{delighat}
\end{align}
and 
that the spacetime is asymptotically 
$\hat{g}_{zz} \rightarrow 1/z^2, \;\; \hat{g}_{ij} \rightarrow 1/z^2, \;\; \tilde{g}_{ij} \rightarrow 0$ 
as $z \rightarrow 0$.

The gravity action consists of the Einstein-Hilbert action\footnote{The Riemann curvature $R^{\mu}_{{\nu}{\rho}{\sigma}}$ is defined by
$
  R^{\mu}_{{\nu}{\rho}{\sigma}} = \del_{{\rho}} \Gamma^{\mu}_{{\nu}{\sigma}} - \del_{{\sigma}} \Gamma^{\mu}_{{\nu}{\rho}} + \Gamma^{\mu}_{\lambda{\rho}} \Gamma^{\lambda}_{{\sigma}{\nu}} - \Gamma^{\mu}_{{\nu}{\lambda}} \Gamma^{\lambda}_{{\sigma}{\rho}} $
and the Ricci tensor is defined by
 $ R_{{\mu}{\nu}} = R^{\rho}_{{\mu}{\rho}{\nu}}$. } 
and the Gibbons-Hawking term:
\begin{align}
  S_{g} = \frac{1}{16\pi G_N} \int d^{d+1} x \sqrt{g} \left(- R + 2\Lambda \right) + S_{GH}\ ,
\end{align}
where the bulk cosmological constant $\Lambda$ is given by
$\Lambda = \frac{-d(d-1)}{2}$.
The matter action is given by
\begin{align}
  S_M = \int d^{d+1} x \sqrt{g} \left( \frac{1}{2} \del_{\mu} \Psi \del^{\mu} \Psi + U_1({\Psi}) + \frac{1}{2} \del_{\mu} \Phi \del^{\mu} \Phi + U_2(\Phi) \right) \ ,
\end{align}
where $\Psi$ is a scalar field that gives the background metric $\hat{g}$,
and $\Phi$ is a perturbation that gives a back reaction corresponding to 
$\tilde{g}_{\mu\nu}$ in (\ref{expansion of metric}).
In what follows, we keep the contribution up to the second order in $\Phi$.
%is a dual of the operator $\mathcal{O}$ on the field theory side.
%$V(\Phi),U(\Psi)$

The Einstein equations are given by
\begin{align}
  R_{\mu\nu} - \frac{1}{2} g_{\mu\nu} R + \Lambda g_{\mu\nu} = 8\pi G_N T_{\mu\nu} \ .
\label{Einstein equation}
\end{align}
We substitute (\ref{expansion of metric}) into (\ref{Einstein equation}) and expand 
(\ref{Einstein equation}) 
in terms of $\tilde{g}_{\mu\nu}$.
Then, the zeroth and first orders read
\begin{align}
&  \hat{R}_{\mu\nu} - \frac{1}{2} \hat{g}_{\mu\nu} \hat{R} + \Lambda \hat{g}_{\mu\nu} = 8\pi G_N \hat{T}_{\mu\nu}   \ , \label{Eisnstein equation 1}\\
&\tilde{R}_{\mu\nu} - \frac{1}{2} \tilde{R} \hat{g}_{\mu\nu} - \frac{1}{2} \tilde{g}_{\mu\nu} \hat{R} + \Lambda\tilde{g}_{\mu\nu} = 8 \pi G_{N} \tilde{T}_{\mu\nu}  \ , \label{Einstein equation 2}
\end{align}
respectively, where
\begin{align}
&  \hat{T}_{\mu\nu} = \del_{\mu} \Psi \del_{\nu} \Psi - \hat{g}_{\mu\nu} \left( \frac{1}{2} \del_{\rho} \Psi \del^{\rho} \Psi + U_1(\Psi) \right)  \ , \\
&\tilde{T}_{\mu\nu} = \del_{\mu} \Phi \del_{\nu} \Phi 
  - \hat{g}_{\mu\nu} \left( \frac{1}{2} \del_{\rho} \Phi \del^{\rho} \Phi + \frac{1}{2}U_2''(0) \Phi^2  \right) 
  -  \tilde{g}_{\mu\nu} \left( \frac{1}{2} \del_{\rho} \Psi \del^{\rho} \Psi + U_1(\Psi) \right)
   \ 
\end{align}
with $U_2''(\Phi) = \del^2 U_2 (\Phi) / \del \Phi^2$.

Because of (\ref{delighat}), there exist $d$ Killing vectors corresponding to translations
in the $(d +1) $-dimensional background spacetime. 
We denote one of those by $\xi^{\mu}$, which satisfies
\begin{align}
  &\hat{\nabla}_{\mu} \xi_{\nu} + \hat{\nabla}_{\nu} \xi_{\mu} = 0 \ , \nonumber\\
  &\mathcal{L}_{\xi} \Psi = 0 \ . 
\label{Killing equation}
\end{align}
We can further assume that the invariance under the  translation corresponding to $\xi^{\mu}$
is preserved by the perturbation:
\begin{align}
  &\mathcal{L}_{\xi} \tilde{g}_{\mu\nu} = 0, \;\; \mathcal{L}_{\xi}\Phi = 0 \  . 
  \label{lie derivative}
\end{align}

We contract $\xi^{\mu}$ with the Einstein equations (\ref{Einstein equation 2}) as
\begin{align}
  \frac{\xi^{\mu} \xi^{\nu}}{\xi^2} \left(\tilde{R}_{\mu\nu} - \frac{1}{2} \tilde{R} \hat{g}_{\mu\nu} - \frac{1}{2} \tilde{g}_{\mu\nu} \hat{R} + \Lambda\tilde{g}_{\mu\nu}\right) = 8 \pi G_{N} \frac{\xi^{\mu} \xi^{\nu}}{\xi^2}\tilde{T}_{\mu\nu}   \ ,
\label{xicontraction}
\end{align}
where $\xi^2 = \hat{g}_{\mu\nu} \xi^{\mu} \xi^{\nu}$ and expand 
$\tilde{R}_{\mu\nu}$ to the first order in
$\tilde{g}_{\mu\nu}$ as
\begin{align}
  \tilde{R}_{\mu\nu} = \frac{1}{2} \hat{g}^{\rho\sigma} (\hat{\nabla}_{\rho} \hat{\nabla}_{\mu} \tilde{g}_{\nu \sigma} + \hat{\nabla}_{\rho} \hat{\nabla}_{\nu} \tilde{g}_{\mu \sigma} - \hat{\nabla}_{\rho} \hat{\nabla}_{\sigma} \tilde{g}_{\mu \nu} - \hat{\nabla}_{\nu} \hat{\nabla}_{\mu} \tilde{g}_{\rho \sigma}  )\ .
\label{Rtilde}
\end{align}
By using (\ref{Eisnstein equation 1}), (\ref{Killing equation}), (\ref{lie derivative}) 
and (\ref{Rtilde}),  we  obtain from  (\ref{xicontraction}) 
\begin{align}
  &\frac{1}{2} \hat{\nabla}_{\rho} 
  \left\{ \frac{1}{\xi^2} \hat{\nabla}^{\mu}(\xi^2) \hat{g}^{\rho\sigma} \tilde{g}_{\mu\sigma}
  -\hat{\nabla}^{\rho} \left(\frac{\xi^{\mu} \xi^{\nu}}{\xi^2} \tilde{g}_{\mu\nu}\right) 
  -\frac{\xi^{\rho} \xi^{\sigma}}{\xi^2} \hat{g}^{\mu\nu} \hat{\nabla}_{\sigma} \tilde{g}_{\mu\nu} 
  + \frac{2}{\xi^2} \hat{\nabla}^{\rho} (\xi^{\mu} \xi^{\nu}) \tilde{g}_{\mu\nu} \right. \n
  &\left.
   \ \ \ \ \ \ \ \ 
  + \hat{\nabla}^{\rho} \left(\frac{2}{\xi^2}\right) \xi^{\mu} \xi^{\nu} \tilde{g}_{\mu\nu} 
  - \frac{1}{2}\frac{1}{\xi^2} \hat{\nabla}^{\rho}(\xi^2) \hat{g}^{\mu\nu} \tilde{g}_{\mu\nu}
  \right\}
  + \frac{1}{2} \left\{ \hat{g}^{\mu\nu} \hat{\nabla}_{\rho} \hat{\nabla}^{\rho} \tilde{g}_{\mu\nu} 
    - \hat{g}^{\mu\nu} \hat{g}^{\rho\sigma} \hat{\nabla}_{\rho} \hat{\nabla}_{\nu} \tilde{g}_{\mu\sigma} \right\}\n
  &- \frac{1}{2 } \hat{\nabla}^{\rho} \left(\frac{1}{\xi^2} \right) \hat{\nabla}^{\mu} (\xi^2) \tilde{g}_{\mu\rho} 
  - \hat{\nabla}^{\rho} \left( \frac{1}{\xi^2} \right) \hat{\nabla}_{\rho}(\xi^{\mu} \xi^{\nu}) \tilde{g}_{\mu\nu}
  + \hat{\nabla}_{\rho} \left(\frac{1}{\xi^2}\right) \hat{\nabla}\xi^2 \frac{\xi^{\mu} \xi^{\nu}}{\xi^2} \tilde{g}_{\mu\nu} \n
  &+ \frac{1}{\xi^2} \hat{\nabla}_{\rho} \xi^{\sigma} \hat{\nabla}^{\rho} \xi_{\sigma} \frac{\xi^{\mu} \xi^{\nu}}{\xi^2} \tilde{g}_{\mu\nu} 
  - \frac{2}{\xi^2} \hat{\nabla}_{\rho} \xi^{\mu} \hat{\nabla}^{\rho} \xi^{\nu} \tilde{g}_{\mu\nu} 
  + \frac{1}{4} \left\{\hat{\nabla}_{\rho} \left(\frac{1}{\xi^2}\right) \hat{\nabla}^{\rho}  - \frac{2}{\xi^2} \hat{\nabla}_{\rho} \xi^{\sigma} \hat{\nabla}_{\sigma} \xi^{\rho} \right\} \hat{g}^{\mu\nu} \tilde{g}_{\mu\nu}\n
  &= - 4\pi G_N \left(\del_{\mu} \Phi \del^{\mu} \Phi +  U_2''(0)\Phi^2 \right) \ .
  \label{xixi tilde g}
\end{align}

By using the equation of motion for $\Phi$, we see that the RHS of (\ref{xixi tilde g}) is a total derivative term, $-4\pi G_N \hat{\nabla}_{\mu} \left(\Phi \hat{\nabla}^{\mu} \Phi \right) $.
The third and fourth lines in the LHS are calculated as
\begin{align}
  &\frac{1}{2}\xi^{\alpha} \del_z \left( \frac{\hat{g}_{\alpha\beta}}{\xi^2} \right)
  \left\{ \frac{1}{\xi^2} \del^z\xi^2 \hat{g}^{\nu\beta} \xi^{\mu} \tilde{g}_{\mu\nu} 
  +  \del^z \hat{g}_{\lambda\sigma} \hat{g}^{\beta\sigma} \xi^{\lambda} \frac{\xi^{\mu} \xi^{\nu}}{\xi^2} \tilde{g}_{\mu\nu}
  - \del^z \hat{g}_{\lambda\sigma} \hat{g}^{\mu\beta} \hat{g}^{\nu\sigma} \xi^{\lambda} \tilde{g}_{\mu\nu}  \right.\n
  &\left. \ \ \ \ \ \ \ \ \ \ \ \ \ \ \ \ \ \ \ \ \ \ 
   + \frac{1}{2} \del^z \hat{g}_{\lambda\sigma} \hat{g}^{\beta\sigma} \xi^{\lambda} \hat{g}^{\mu\nu} \tilde{g}_{\mu\nu}
  \right\} \ .
\end{align}
Then, in order for the LHS to be total derivative terms, the above expression must vanish.
This leads us to impose a condition
\begin{align}
  \xi^{\mu} \del_{z} \left(\frac{\hat{g}_{\mu\nu}}{\xi^2}\right) = 0 \ .
  \label{hatg condition}
\end{align}
Thus, (\ref{xixi tilde g}) reduces to
\begin{align}
&  \frac{1}{2} \hat{\nabla}_{\rho} 
  \left\{ \frac{1}{\xi^2} \hat{\nabla}^{\mu}(\xi^2) \hat{g}^{\rho\sigma} \tilde{g}_{\mu\sigma}
  -\hat{\nabla}^{\rho} \left(\frac{\xi^{\mu} \xi^{\nu}}{\xi^2} \tilde{g}_{\mu\nu}\right) 
  -\frac{\xi^{\rho} \xi^{\sigma}}{\xi^2} \hat{g}^{\mu\nu} \hat{\nabla}_{\sigma} \tilde{g}_{\mu\nu} 
  + \frac{2}{\xi^2} \hat{\nabla}^{\rho} (\xi^{\mu} \xi^{\nu}) \tilde{g}_{\mu\nu} \right. \n
&  \left.
   \ \ \ \ \ \ \ \ 
  + \hat{\nabla}^{\rho} \left(\frac{2}{\xi^2}\right) \xi^{\mu} \xi^{\nu} \tilde{g}_{\mu\nu} 
  - \frac{1}{2}\frac{1}{\xi^2} \hat{\nabla}^{\rho}(\xi^2) \hat{g}^{\mu\nu} \tilde{g}_{\mu\nu}
  + \hat{g}^{\mu\nu} \hat{\nabla}^{\rho} \tilde{g}_{\mu\nu} 
    - \hat{g}^{\mu\nu} \hat{g}^{\rho\sigma}  \hat{\nabla}_{\nu} \tilde{g}_{\mu\sigma} \right\}\n
&  = - 4\pi G_N \hat{\nabla}_{\mu} \left(\Phi \hat{\nabla}^{\mu} \Phi \right)  \ .
\label{xicontraction2}
\end{align}
We assume that the boundary where the dual field theory lives is specified
by $z=\epsilon$. Then, integrating both sides of (\ref{xicontraction2}) over the bulk yields
\begin{align}
&   \frac{1}{2} \int d^d x \sqrt{\hat{g}}  
    \left\{ \hat{g}^{zz} \del_{z}\left( \left(\hat{g}^{ij} - \frac{\xi^{i}\xi^{j}}{\xi^2} \right) \tilde{g}_{ij}\right) 
    - \frac{\hat{g}^{zz}}{2} \left( \del_z \hat{g}^{ij} + \frac{1}{\xi^2} \del_z \xi^2 \hat{g}^{ij} \right) \tilde{g}_{ij}\right\}\n
&  =  8 \pi G_N S_{\Phi  \; \rm{on-shell}}   \ ,
  \label{int einstein}
\end{align}
where (\ref{metric}) is used, and the on-shell action for $\Phi$ in the RHS is given by
\begin{align}
  S_{\Phi \; \rm{on-shell}} = - \frac{1}{2} \int_{z = \epsilon} d^d x \sqrt{\hat{g}} \hat{g}^{zz} \Phi \del_{z} \Phi  \ .
\end{align}

Hereafter, we identify the direction of $\xi^i$ with that of $x^0$ such that $\xi^z=0, \; \xi^0=1$
and $\xi^a=0$.
%Note that the following can be calculated without generality.
%We introduce the coordinates $\sigma^a$ $(a=1,\ldots,d-1)$ that parametrize the space 
%pependicular to $\xi^i$ in $d$ dimensions
We make an ADM-like decomposition of the metric $g_{ij}$ in $d$ dimensions as
\begin{align}
  g_{ij} = \left(
    \begin{array}{cc}
      \frac{1}{N^2} & -\frac{N_b}{N^2}  \\
      -\frac{N_a}{N^2} &  \gamma_{ab} + \frac{N_a N_b}{N^2} 
    \end{array}
  \right) \ ,  \;\;
  g^{ij} = \left(
    \begin{array}{cc}
      N^2 + N^a N_a & N^b  \\
      N^a &  \gamma^{ab}
    \end{array}
  \right)  \ ,
\end{align}
where $a,b=1,\ldots,d-1$.
%and $\gamma_{ab}$ is the induced metric of the space perpendicular to $\xi^i$.
Expanding $N$, $N_a$ and $\gamma_{ab}$ around  the background as
$N=\hat{N} + \tilde{N}$, $N_a=\hat{N}_a + \tilde{N}_a$ and 
$\gamma_{ab}=\hat{\gamma}_{ab} + \tilde{\gamma}_{ab}$, leads to
\begin{align}
  \hat{g}_{ij} = \left(
    \begin{array}{cc}
      \frac{1}{\hat{N}^2} & -\frac{\hat{N}_b}{N^2}  \\
      -\frac{\hat{N}_a}{\hat{N}^2} &  \hat{\gamma}_{ab} + \frac{\hat{N}_a \hat{N}_b}{\hat{N}^2} 
    \end{array}
  \right), \ 
  \tilde{g}_{ij} = \left(
    \begin{array}{cc}
      -2\frac{\tilde{N}}{\hat{N}^3} & 2\frac{\hat{N}_b \tilde{N}}{\hat{N}^3} - \frac{\tilde{N}_b}{\hat{N}^2} \\
      2\frac{\hat{N}_a \tilde{N}}{\hat{N}^3} - \frac{\tilde{N}_a}{\hat{N}^2} &  \tilde{\gamma}_{ab} -2\frac{\hat{N}_a \hat{N}_b}{\hat{N}^2} \tilde{N} + \frac{\tilde{N}_a \hat{N}_b}{\hat{N}^2} + \frac{\hat{N}_a \tilde{N}_b}{\hat{N}^2} 
    \end{array}
  \right)  \ .
\end{align}

Note that $\xi^2$ and $\sqrt{\hat{g}}$ are given by
\begin{align}
  \xi^2 = \frac{1}{\hat{N}^2}
  ,\ \sqrt{\hat{g}} = \frac{\sqrt{\hat{N}^2 + \hat{N}^a\hat{N}_a}}{\hat{N}^2} \sqrt{\hat{g}_{zz}}\sqrt{\hat{\gamma}}
\label{xi^2 and sqrtg}
\end{align}
and  that (\ref{hatg condition}) is equivalent to
\begin{align}
  \del_z \hat{N}_a = 0 \ .
  \label{hatNa condition}
\end{align}

By substituting (\ref{xi^2 and sqrtg}) and (\ref{hatNa condition}) into (\ref{int einstein}), 
we obtain
\begin{align}
  8 \pi G_N S_{\Phi\; \rm{on-shell} } 
  =\frac{1}{2}\int_{z = \epsilon} d^d x \frac{\sqrt{\hat{N}^2 + \hat{N}^a \hat{N}_a}}{\hat{N}^2\sqrt{\hat{g}_{zz}}} \sqrt{\hat{\gamma}} 
   \left\{ \del_{z}\left( \hat{\gamma}^{ab} \tilde{\gamma}_{ab}\right) 
   - \frac{1}{2} \left( \del_z \hat{\gamma}^{ab} - \frac{1}{\hat{N}^2} \del_z \hat{N}^2 \hat{\gamma}^{ab} \right) \tilde{\gamma}_{ab}\right\}   \ .
  \label{int einstein 2}
\end{align}

The boundary specified by $z=\epsilon$ where the dual field theory lives 
is a codimension-2 hyperplane perpendicular to $\xi^i$.
The volume of hyperplane is given up to the first order in $\tilde{\gamma}_{ab}$ as
\begin{align}
  V &= \int_{z = \epsilon} d^{d-1} x \sqrt{\gamma}\n
  &= \int_{z = \epsilon} d^{d-1} x \sqrt{\hat{\gamma}} (1 + \frac{1}{2}\hat{\gamma}^{ab} \tilde{\gamma}_{ab}) \ .
\end{align}
We subtract the zeroth order contribution from this and obtain
\begin{align}
\delta V= \frac{1}{2} \int_{z = \epsilon} d^{d-1} x \sqrt{\hat{\gamma}} 
\hat{\gamma}^{ab} \tilde{\gamma}_{ab}  \ .
\end{align}

We consider the $z$ derivative of $\delta V $:
\begin{align}
  \delta V ' &= \frac{1}{2}\int_{z = \epsilon} d^{d-1} x \del_z (\sqrt{\hat{\gamma}}) \ \hat{\gamma}^{ab} \tilde{\gamma}_{ab} + \delta {v}' \ ,  \label{delta V prime} \\
  &\delta v'= \frac{1}{2}\int_{z = \epsilon} d^{d-1} x \sqrt{\hat{\gamma}} \ \del_z \left(\hat{\gamma}^{ab} \tilde{\gamma}_{ab}\right) \ ,
\end{align}
where the prime represents the $z$-derivative.
Here the first term and the second term in the RHS of (\ref{delta V prime}) represent
the canonical scaling contribution and a nontrivial scaling contribution, respectively.
Then , (\ref{int einstein 2}) is rewritten as
\begin{align}
  8 \pi G_N S_{\Phi \; \rm{on-shell}} 
  =\int d \tau \frac{\sqrt{\hat{N}^2 + \hat{N}^a \hat{N}_a}}{\hat{N}^2\sqrt{\hat{g}_{zz}}} 
   \left\{ \delta v'
   - \frac{1}{4}\int_{z = \epsilon} d^{d-1}x \sqrt{\hat{\gamma}}\left( \del_z \hat{\gamma}^{ab} - \frac{1}{\hat{N}^2} \del_z \hat{N}^2 \hat{\gamma}^{ab} \right) \tilde{\gamma}_{ab}\right\}  \; .
   \label{Sonshell and dV}
\end{align}
Here we require the second term in the RHS of  (\ref{Sonshell and dV}) to vanish.
Then, $\hat{\gamma}_{ab}$ is determined as
\begin{align}
  \hat{\gamma}_{ab} = \frac{1}{\hat{N}^2} \ C_{ab}
\end{align}
with $C_{ab}$ being a constant tensor.
Furthermore, (\ref{hatNa condition}) implies that $\hat{g}_{ij}$ is expressed as
\begin{align}
  \hat{g}_{ij} 
%  &= \frac{1}{\hat{N}^2}\left(
%    \begin{array}{cc}
%      1 & -\hat{N}_b  \\
%      -\hat{N}_a &  C_{ab} + \hat{N}_a \hat{N}_b
%    \end{array}
%  \right)\n
    = \frac{1}{\hat{N}^2} D_{ij} 
  \label{sol backgraund metric}
\end{align}
with  $D_{ij}$ being a constant tensor.

$D_{ij}$ is diagonalizable so that $\hat{N}_a$ can be set to zero and we redefine $g_{zz} \rightarrow \frac{1}{z^2}, g_{ij} \rightarrow \frac{g_{ij}}{z^2}$.
Thus, from  (\ref{Sonshell and dV}), we obtain
\begin{align}
  S_{\Phi \; \rm{on-shell}} = \frac{T}{8\pi G_N} \delta v' \ ,
  \label{formula}
\end{align}
where we have used
\begin{align}
  \frac{1}{\hat{N}} = \frac{\hat{g}_{\tau\tau}}{z} 
\end{align}
and
\begin{align}
  T = \int d \tau \hat{g}_{\tau\tau}  \ .
\end{align}

The AdS metric satisfies (\ref{sol backgraund metric}) so that we can take
\begin{align}
  \hat{N}=z, \hat{N}_a = 0, \hat{\gamma}_{ab} = \frac{1}{z^2}\delta_{ab} \ ,
\end{align}
and we reproduce the result in \cite{Tsuchiya:2020drh}.

(\ref{sol backgraund metric}) indicates that the background spacetime has
$d$-dimensional Poincare invariance.
This implies that the original dual field theory also has it.
%the time direction $x^0$ and the space directions $x^a$ are related such that
%the field theory on the boundary must have the Poincare symmetry.

The background spacetime $\hat{g}_{\mu\nu}(z)$ satisfying (\ref{sol backgraund metric}) 
and the background matter field $\Psi(z)$
are determined by the equations of motion
\begin{align}
&  (d-1) \left(\frac{f''}{f} - \frac{f'^2}{f^2} + \frac{1}{z} \frac{f'}{f}\right)  = - 16\pi G_N \del_z \Psi \del_z \Psi
  \ , 
  \label{background E eq}  \\
&  z^2\Psi'' -(d-1)z\Psi' + \frac{d}{2}\frac{f'}{f}z^2 \Psi' = \frac{\del U_1(\Psi)}{\del \Psi}   \ ,
  \label{background f eq}
\end{align}
where 
%$'=\del_z$ and  
we define $\frac{1}{\hat{N}^2} = \frac{f(z)}{z^2}$.

As an example of solutions, we consider the GPPZ flow \cite{Girardello:1999bd} which is dual to
4-dimensional $\mathcal{N}=1$ super Yang-Mills theory, where 
the scalar field has a non-zero VEV. In this case, the potential $U_1(\Psi)$ is given by
\begin{align}
  U_1(\Psi) = -\frac{3}{32\pi G_N} \left[- 5 + \left(\cosh \left(\frac{4\sqrt{\pi G_N}}{\sqrt{3}}\Psi\right)\right)^2 + \cosh \left(\frac{4\sqrt{\pi G_N}}{\sqrt{3}}\Psi\right)\right] \ ,
\end{align}
and the solution to (\ref{background E eq}) and (\ref{background f eq}) is given by
\begin{align}
  \Psi = \frac{\sqrt{3}}{4\sqrt{\pi G_N}} \log \frac{1+ z}{1-z} , \ \ \ \ \ f(z) = 1-z^2  \ .
\end{align}

%%%%%%%%%%%%%%%%%%%%%%%%%%%%%%%%%%%%%%%%%%%%%%%%%%%%%%%%%%%%%%%%%%%%%
\section{Information metric for a dual operator to bulk scalar field}
\setcounter{equation}{0}
%%%%%%%%%%%%%%%%%%%%%%%%%%%%%%%%%%%%%%%%%%%%%%%%%%%%%%%%%%%%%%%%%%%%%

In this section, we
introduce the quantum information metric 
and show that the one for the original and perturbed theories on the field theory side
is represented by the geometrical quantity $\delta v$ in (\ref{formula}).

We consider a field theory defined by a Lagrangian density ${\cal L}_0$ on $d$-dimensional 
Euclidean spacetime whose
coordinates are $x^i$ $(i=0,\ldots,d-1)$, where $x^0 \equiv \tau$ is viewed as the Euclidean time.
We also consider another field theory with a Lagrangian density ${\cal L}$  obtained 
by perturbing the theory as
\begin{align}
  \mathcal{L} = \mathcal{L}_0 + \phi_{(0)}(\vec{x}) \mathcal{O}(x) \ ,
\end{align}
where $\mathcal{O}(x)$ is a scalar operator and $\phi_{(0)}(\vec{x})$ is a source independent of the time $\tau$. 
%Note that we should choose the independed variable acounding to the lie derivative (\ref{lie %derivative}) of the bulk field with a killing vector. 

We denote the ground states of the theories ${\cal L}_0$ and ${\cal L}$ by $\ket{\Omega_0}$ and
$\ket{\Omega}$, respectively. Then, the inner product $\braket{\Omega|\Omega_0}$ is 
given by a path integration
\begin{align}
  \braket{\Omega|\Omega_0} = (ZZ_0)^{-1/2} \int \mathcal{D}\phi \ \exp
  \left[-\int d^{d-1} x (\int_{-\infty}^0 d\tau \mathcal{L}_0 + \int^{\infty}_0  d\tau \mathcal{L})\right] \ ,
\end{align}
where $Z_0$ and $Z$ is the partition functions of the theories $\mathcal{L}_0$ 
and $\mathcal{L}$, respectively.
We assume that $\braket{\mathcal{O}(x)}_0=0$
and the time reversal symmetry:
$\braket{\mathcal{O}(\tau,\vec{x}) \mathcal{O}(\tau',\vec{x}')}_0 = \braket{\mathcal{O}(-\tau,\vec{x}) \mathcal{O}(-\tau',\vec{x}')}_0$, where
\begin{align}
\braket{\cdots}_0 =\frac{1}{Z_0}\int \mathcal{D}\phi \cdots \exp\left[ -\int d^dx  \ 
\mathcal{L}_0 \right] \ .
\end{align}
The information metric $\mathcal{G}$ that measures the distance between the ground states of the two theories
is obtained by expanding $\braket{\Omega|\Omega_0}$ up to the second order in $\phi_{(0)}$:
\begin{align}
\mathcal{G}=  \frac{1}{T}(1 - \braket{\Omega'|\Omega}) &= \frac{1}{2T} \int_0^{\infty} d\tau \int_{-\infty}^0  d\tau' \int d^{d-1} x \int d^{d-1} x' \braket{\delta \mathcal{L}(\tau,x) \delta \mathcal{L}(\tau',x') }_0 \n
   &=  \frac{1}{2T} \int_0^{\infty} d\tau \int_{-\infty}^0  d\tau' \int d^{d-1} x \int d^{d-1} x' \phi_{(0)}(\vec{x}) \braket{\mathcal{O}(\tau,\vec{x}) \mathcal{O}(\tau',\vec{x}')}_0 \phi_{(0)}(\vec{x}') \ ,
   \label{eq: information metric of scalar}
\end{align}
where $T$ is the volume of the time direction. Here we assume that we make an appropriate regularization for the two point function of ${\cal O}$ to
suppress a divergence occurring at $\tau = \tau' =0$. In the case in which ${\cal L} _0$ is a CFT with ${\cal O}$ a primary operator, a regularization is given in section 3 of 
\cite{Tsuchiya:2020drh}.

We apply the above result to the case in the previous section: 
the theory $\mathcal{L}_0$ possesses a gravity dual corresponding to the background geometry with the $d$-dimensional Poincare invariance, 
and 
the operator $\mathcal{O}$ corresponds to  a scalar field $\Phi$,
which coincides with $\phi_{(0)}$ on the boundary.
The $\tau$-independence of $\phi_{(0)}$ is consistent with (\ref{lie derivative}).
We consider a situation where the classical approximation on the gravity side is valid.
By using the GKP-Witten relation
\begin{align}
  \left\langle \exp \left[-\int d^d x \phi_{(0)}(x) \mathcal{O}(x) \right] \right\rangle
   = \exp \{-S_{\Phi \; \rm{on-shell}}\}
\label{GKP-Witten}\ ,
\end{align}
we can show that
\begin{align}
S_{\Phi \; \rm{on-shell}} = -4T\mathcal{G} \ .
\end{align}
Thus, by using (\ref{formula}), we obtain a formula,
\begin{align}
  \mathcal{G} = -\frac{1}{32\pi G_N} \delta v' \ . 
\end{align}
This formula was obtained in \cite{Tsuchiya:2020drh} in the case where
a CFT is perturbed by a primary operator ${\cal O}$ on the field theory side and the
background geometry on the gravity side is given by the AdS.

%%%%%%%%%%%%%%%%%%%%%%%%%%%%%%%%%%%%%%%%%%%%%%%%%%%%%%%%%%%%%%%%%%%%%
\section{Conclusion}
\setcounter{equation}{0}
%%%%%%%%%%%%%%%%%%%%%%%%%%%%%%%%%%%%%%%%%%%%%%%%%%%%%%%%%%%%%%%%%%%%%
In this letter, we considered a field theory that has a gravity dual, and perturbed it by an operator
which corresponds to a scalar field in the bulk. We performed a covariant calculation to find the condition
that must be satisfied by the bulk geometry in order that the on-shell action for the scalar field is represented by a back reaction to the volume of a codimension-2 surface.
The condition implies the Poincare invariance of the original field theory.
We saw that the quantum information metric that measures the distance
between the ground states of the original and perturbed theories
is represented by the on-shell action. While we considered only a perturbation by a scalar field,
we should obtain the same results for perturbations by vector and tensor fields.
Thus, we conclude that the universal formula in \cite{Tsuchiya:2020drh} 
that represents the quantum information metric
in terms of the back reaction to the volume of the codimension-2 surface is extended
to the case of 
a general gauge/gravity correspondence if the
above condition is satisfied.
It is interesting to elucidate what is represented by the extra terms in the RHS of (\ref{Sonshell and dV}) which we have if we do not impose the condition of the Poincare invariance.  
We hope that our result
leads us to gain deeper understanding
of the relationship between quantum information and quantum geometry.

%%%%%%%%%%%%%%%%%%%%%%%%%%%%%%%%%%%%%%%%%%%%%%%%%%%%%%%%%%%%%%%%%%%%%%%%
\section*{Acknowledgments}
A.T. was supported in part by Grant-in-Aid for Scientific Research (No. 18K03614 and No. 21K03532) from
Japan Society for the Promotion of Science.
K.Y. was supported in part by Grant-in-Aid for JSPS
Fellows (No. 20J13836).

%%%%%%%%%%%%%%%%%%%%%%%%%%%%%%%%%%%%%%%%%%%%%%%%%%%%%%%%%%%%%%%%%%%%%%%%%
%               bibliography
%%%%%%%%%%%%%%%%%%%%%%%%%%%%%%%%%%%%%%%%%%%%%%%%%%%%%%%%%%%%%%%%%%%%%%%%%

    %%%%%%%%%%%%%%%%%%%%%%%%%%%%%%%%%%%%%%%%%%%%%%%%%%%%%%%%%%%%%%%%%%%%%%%%
    
    \end{document}